\documentclass[fleqn,usenatbib]{mnras}
\usepackage{amsmath}
\usepackage{graphicx}
\usepackage{amssymb,txfonts}

\newcommand{\msun}{{\rm M}_{\sun}}
\newcommand{\rsun}{{\rm R}_{\sun}}
\newcommand{\lsun}{{\rm L}_{\sun}}
\newcommand{\source}{V404 Cyg}
\newcommand{\integral}{{\textit{INTEGRAL}}}

\newcommand{\ginga}{{\textit{Ginga}}}

\newcommand{\swift}{{\textit{Neil Gehrels Swift}}}

\title[Non-conservative mass transfer in V404 Cyg]{Non-conservative mass transfer in stellar evolution and the case of V404 Cyg/GS 2023+338}
\author[J. Zi\'o{\l}kowski and A. A. Zdziarski]{Janusz Zi\'o{\l}kowski\thanks{E-mail:
jz@camk.edu.pl, aaz@camk.edu.pl} and Andrzej A. Zdziarski\footnotemark[1]\\
Nicolaus Copernicus Astronomical Center, Polish Academy of Sciences, Bartycka 18, PL-00-716 Warszawa, Poland}

\begin{document}

\date{Accepted 2018 July 18. Received 2018 July 17; in original form 2018 April 19}

\pagerange{\pageref{firstpage}--\pageref{lastpage}} 
\pubyear{2018}

\maketitle

\label{firstpage}

\begin{abstract}
We consider donor evolution and mass transfer in the microquasar V404 Cyg/GS 2023+338. Based on X-ray observations of its two outbursts, its average mass accretion rate is substantially lower than our model mass-loss rate from its low-mass giant donor. A likely solution to this discrepancy is that a large fraction of the mass flowing from the donor leaves the binary in the form of outflows from the accretion disc around the accretor. The outflow can be parameterizes by the fractions of the mass and angular momentum leaving the system. We calculate the latter as a function of the radius of the accretion disc from which the outflow takes place. We then perform detailed evolutionary calculations for V404 Cyg. Given our estimated average accretion rate, $>$70 per cent of the mass lost from the donor has to leave the binary. The allowed solutions for the actual donor mass loss rate are parameterized by our two outflow parameters. Our results are in agreement with the observed outflows from the outer disc as well as with the variable near-Eddington accretion observed during the outbursts, compatible with outflows from the vicinity of the black hole. We then calculate the expected rate of the orbital period change. Its future measurements can constrain the presence of an outflow in V404 Cyg and its parameters.
\end{abstract}

\section{Introduction}
\label{intro}

GS 2023+338 was discovered by \ginga\/ in May 1989 as a new X-ray transient \citep{makino89}. Its optical counterpart was immediately identified as a variable star V404 Cyg \citep{wagner89}. V404 Cyg was originally found as the optical counterpart of Nova Cyg 1938, which was classified as a classical nova. However, \citet{charles89} found that it was neither classical nor recurrent nova but a low mass X-ray binary. \citet*{casares92} discovered absorption lines of an early K star and determined the orbital period as $P\simeq 6.473 \pm 0.001$ d. The radial velocity amplitude gave the mass function of $6.08 \pm 0.06 \msun$, implying the presence of a black hole in the system \citep{cc94}. Those authors also obtained a relatively precise estimate of the rotational broadening of the absorption lines of the donor of $v \sin i \simeq 39.1 \pm 1.2$ km s$^{-1}$, which implies the mass ratio of $q \equiv M_2/M_1 \simeq 0.060^{+0.004}_{-0.005}$, where $M_1$ and $M_2$ are the masses of the black hole and the donor, respectively. \citet{miller_jones09} measured the radio parallax of the system and found the distance of $D\simeq 2.39 \pm 0.14$ kpc. \citet*{khargharia10} presented near-infrared spectroscopy of the donor component. They established its spectral type as K3 III with the uncertainty of one subtype, i.e., within K2--K4 III, i.e., the star is a giant, confirming \citet{king93}. Then they modeled the H-band light curve and on the basis of their fit determined the inclination of the system as $i \simeq 67^{+3}_{-1}\degr$. Using this value, the mass function and the mass ratio, the mass of the black-hole and donor component is $M_1 \simeq 9.0^{+0.2}_{-0.6} \msun$ and $M_2\simeq 0.54 \pm 0.05 \msun$, respectively. 

The above observational data permit us to estimate relatively precisely the radius and the luminosity of the donor. Using the expression for the Roche lobe radius of \citet{paczynski67} together with the third Kepler law, we obtain the donor radius of $R_2 \simeq 5.50_{-0.18}^{+0.17} \rsun$. The luminosity can be estimated from the donor spectral type. The calibration of \citet{cox00} for K3$^{+1}_{-1}$III \citep{khargharia10} implies the effective temperature of $T_2=4274^{+116}_{-113}$ K. With those $R_2$ and $T_2$, we find the donor luminosity of $L_2\simeq 8.9^{+1.7}_{-1.4}\lsun$. 

In July 2015, \source\/ underwent renewed X-ray and optical activity after 26 years of dormancy \citep{barthelmy15}. Activity was strong (approaching Eddington limit) but lasted only about two weeks \citep{rodriguez15,kimura16,motta17a,motta17b,sanchez17}. After several months of quiescence, the system erupted in December 2015, again for about two weeks but at much lower level of activity \citep*{marti16,motta16}. 

The average accretion rate implied by the X-ray luminosity observed during the 2015 outburst is difficult to estimate due to strong absorption. Fig.\ 4 of \citet{kimura16} gives the bolometric light curve based on hard X-ray observations by \swift/BAT and \integral/ISGRI detectors, where their 15--50 keV and 25--60 keV luminosities (presumed to be weakly affected by absorption) are multiplied by the rather large estimated bolometric-correction factors of 7 and 10, respectively. Even with those corrections, the light curve only occasionally exceeds the Eddington limit, $L_{\rm Edd}$, by a factor $\la$2, and the average luminosity during the two-week outburst is $\sim 0.1 L_{\rm Edd}$ or so. Assuming a 10 per cent efficiency and averaging over the 26 years interval between the two most recent outbursts, we obtain the average accretion rate of $\langle\dot M_1\rangle\simeq 3.5 \times 10^{-11}\msun$ yr$^{-1}$. 

However, it is in principle possible that the strong variability observed in \source\ was due to almost full obscuration of the accretion flow during the low-flux periods \citep{motta17a}. This would increase the average luminosity during the 2015 outburst to $\sim\! L_{\rm Edd}$, yielding $\langle\dot M_1\rangle\simeq 3.5 \times 10^{-10}\msun$ yr$^{-1}$. However, a strong argument against the emission being continuously very bright but occasionally absorbed in the line of sight is the observation of about five bright rings in soft X-rays, caused by dust scattering of the central emission \citep{beardmore16}. Those rings appear to correspond to the main X-ray flares, and show that the intermittent emission was seen not only in our line of sight to the source but also at most other directions. This strongly argues against the intermittency being caused solely by absorbing clouds moving through the line of sight.

We can compare our values with that for the 1989 outburst. \citet*{zycki99} estimated the mass accreted during it as $\simeq 2.6\times 10^{25}$ g (at an assumed 10 per cent efficiency and using the current value of $D=2.39$ kpc), which implies $\langle\dot M_1\rangle\simeq 4.0 \times 10^{-10}\msun$ yr$^{-1}$ when averaged over 33 yr between 1989 and the previous outburst in 1956 (\citealt*{chen97} and references therein). \citet{chen97} also estimated the total radiated energy by integrating over an exponential rise and a decay of the light curve normalized to the peak luminosity for the 1989 outburst. At 10 per cent efficiency, they obtained the accreted mass of $\simeq 2.7\times 10^{25}$ g (for $D=2.39$ kpc), implying an almost the same average $\langle\dot M_1\rangle$ as that of \citet{zycki99}. Taking into account all of the above estimates, $4.0 \times 10^{-10}\msun$ yr$^{-1}$ appears to be a likely upper limit on $\langle\dot M_1\rangle$.

On the other hand, all of those estimates are lower than the value of the donor mass loss rate implied by considering evolution of giants. In particular, an approximate formula of \citet*{webbink83} given by their equation (25a) yields $-\dot M_2\simeq 1.1 \times 10^{-9} \msun$ yr$^{-1}$ for the case of \source. We confirm their estimate by our updated evolutionary model.

This discrepancy can be accounted for by assuming that a fraction of the mass lost from the donor can leave the binary in the form of an outflow. In the case of \source, strong outflows from outer parts of the accretion disc have been observed \citep{munoz_darias16}. Also, outflows from inner disc parts are expected at accretion rates corresponding to $L\ga L_{\rm Edd}$ (e.g., \citealt{poutanen07,sadowski15}). This possibility in \source\ has been studied by \citet{motta17b}. Strong outflows are also possible at low accretion rates, e.g., \citet{yuan14}.

Binary evolution with non-conservative mass transfer, i.e., with an outflow carrying away a fraction of the transferred mass from the system, has been extensively studied in the past. It appears that it was first considered by \citet{pz67}, who parameterized it by the fractions of the mass and angular momentum lost in the outflow. The angular momentum of a binary can be also lost by gravitational radiation, which effect on a close binary was first considered by \citet{paczynski67}, and by magnetic breaking (e.g., \citealt{verbunt93}). The mass loss and emission of gravitational radiation were considered by \citet*{rappaport82}, who, in particular, derived a formula describing changes of the Roche lobe radius in the presence of a non-conservative mass transfer. Other studies include \citet{ziolkowski85}, \citet{tout91} and a review of \citet{verbunt93}. More recently, non-conservative evolution of binary systems has been considered as an explanation of large rates of the orbital period changes seen in a number of low-mass X-ray binaries \citep{disalvo08,burderi09,burderi10,sanna17,iaria18}. Furthermore, \citet{marino17} considered the same effect as an explanation of the average X-ray luminosity of the accreting millisecond pulsar XTE J0929--314 being much lower than that inferred by assuming a conservative evolution driven by gravitational radiation. The fact that the mass transfer in several different kinds of low-mass X-ray binaries appears to be highly non-conservative independently of the nature of the compact object is of significant interest, and probably implies a similar physics of the mass outflows.

In this work, we consider evolution of a binary system in the presence of an outflow. We then present an evolutionary model of the donor in \source, taking into account that effect. 

\section{Evolution of the Roche lobe in the presence of an outflow}
\label{outflow}

Equation (25) of \citet{rappaport82} describes evolution of the Roche lobe radius of the donor during non-conservative mass transfer, i.e., in the presence of mass loss carrying some angular momentum. It is in the form of the partial time derivative of $R_2$, where $R_2$ is the radius of the donor Roche lobe. That equation is parametrized by the fraction of the mass lost by the donor that is accreted onto the accretor, $\beta$, and the specific angular momentum of the mass leaving the system in units of $2\upi a^2/P$, where $a$ is the orbital separation. Here, following \citet{verbunt93}, we instead define the latter parameter ($\alpha$) with respect to the specific angular momentum of the donor measured from the centre of mass, CM. 

The total orbital angular momentum, $J$, of a binary system (in the approximation of two point masses) is the sum of the two angular momenta,
\begin{equation}
J = (M_1 a_1^2+M_2 a_2^2)\Omega =
\left(\frac{G a}{M}\right)^{1/2} M_1 M_2,\quad \Omega=\frac{2\upi}{P}=\sqrt{\frac{GM}{a^3}},
\label{J}
\end{equation}
where $a_1=a M_2/M$ and $a_2=a M_1/M$ are the distances of the accretor and donor, respectively, from the CM, $\Omega$ is the binary angular velocity, $M=M_1+M_2$, and $G$ is the gravitational constant. The total and specific orbital angular momentum of the donor, $J_2$, are
\begin{equation}
J_2 = J\frac{M_1}{M},\quad j_2=J \frac{M_1}{M_2 M}=a_2^2\Omega,
\label{J2}
\end{equation}
respectively. As stated above, we parameterize the outflow by
\begin{align}
&\beta = -\frac{{\rm d}M_1}{{\rm d}M_2}= -\frac{\dot M_1}{\dot M_2}=
1-\frac{{\rm d}M_{\rm esc}}{{\rm d}M_2}, 
\label{f1}\\
&\alpha=\frac{{\rm d}J_{\rm esc}}{{\rm d}J_2}=\frac{j_{\rm esc}}{j_2},
\label{f2}
\end{align}
where $\beta=1$ in a conservative transfer, and $M_{\rm esc}$, $J_{\rm esc}$ and $j_{\rm esc}$ are the mass of the outflowing matter, its angular momentum and its specific orbital angular momentum, respectively. We also use the form of the radius of the Roche lobe for $M_2\la 0.6 M$ as given by
\begin{equation}
R_2\simeq \frac{2 a}{3^{4/3}} \left(\frac{M_2}{M} \right)^{1/3}
\label{roche}
\end{equation}
\citep{paczynski67}. With these definitions, we obtain
\begin{equation}
\frac{{\rm d}\ln R_2}{{\rm d}\ln M_2} = -\frac{5}{3} +2\beta \frac{M_2}{M_1} +
\frac{2}{3}(1-\beta) \frac{M_2}{M} +2 (1-\beta) \alpha \frac{M_1}{M},
\label{rate}
\end{equation}
which is equivalent to equation (25) of \citet{rappaport82} after accounting for their different definition of $\alpha$. It also represents a generalization of equation (12) of \citet{webbink83}, and it reduces to it for $\beta=1$.

We can constrain the possible range of $\alpha$ if we knew from which place the outflow takes place. We first note that the specific angular momentum (with respect to the CM) of matter leaving the donor through $L_1$ is not equal $j_2$. The distance of the $L_1$ point from the CM is $a_{L1}=b_1-a_1$, where $b_1$ is the distance of $L_1$ to the centre of the primary, which is given in Appendix \ref{L1_app}. This yields $\alpha$ at $L_1$ as\footnote{We note that \citet{disalvo08} estimated $\alpha_{L1}$ assuming the $L_1$ point is at the distance $R_2$ from the centre of the donor. This is inaccurate since $R_2$ is the volume-averaged Roche-lobe radius, which is substantially lower than the actual distance of $L_1$ from the centre of the donor.} $\alpha_{L1}= (a_{L1}/a_2)^2$. For $q\equiv M_2/M_1=0.54/9=0.06$, $b_1\simeq 0.76 a$, $a_{L1}\simeq 0.70 a$, and $\alpha_{L1}\simeq 0.55$. Thus, the matter leaving the donor has the specific angular momentum significantly lower than $j_2$.

The radius, $R_{\rm circ}$ (with respect to the centre of $M_1$), at which a particle in a Keplerian orbit around the accretor has the specific angular momentum the same as that at $L_1$ corresponds to the initial disc formation, and it is called the circularization radius \citep*{frank02}. However, the matter falling onto the compact object has to get rid of its angular momentum, which then has to be transferred outward. This angular momentum allows the disc to exist beyond $R_{\rm circ}$, and up to the tidal radius of $R_{\rm tidal}\sim\! 0.9 R_1$ \citep{frank02}, where $R_1$ is the Roche lobe radius of the primary. 

Assuming a Keplerian disc around the accretor in the frame co-rotating with the binary, we can calculate the value of $\alpha$ as a function of the disc radius, $R$. If the outflow takes place from $R$ and carries away its angular momentum, it would remove the $\alpha$ fraction of the specific angular momentum of the donor. The angular momentum of an outflowing particle is the sum of the angular momentum due to the binary motion and that due to the disc rotation. The angular momentum in a given point depends on both $R$ and the azimuthal angle, $\phi$, which we measure with respect to the line connecting the stars. We assume the disc rotation is prograde. The angular momentum is measured with respect the CM, which is at the distance of $a_1$ from the compact object. We denote the distance between the CM and the point of the outflow on the disc as $r$, and the angle between $r$ and $R$ as $\gamma$. We have then a triangle with the sides of $R$, $r$, and $a_1$, and $\phi$ is the angle between $R$ and $a_1$. We can solve the triangle as 
\begin{align}
&r(R,\phi)=R\frac{\sin\phi}{\sin(\phi+\gamma)},\label{rcm}\\
&\gamma(R,\phi)=\frac{\upi-\phi}{2}+\arctan\left( \frac{1-R/a_1}{1+R/a_1}\cot\frac{\phi}{2}\right).\label{angle}
\end{align}
The velocity of the particle is the sum of the binary and disc velocities, and its component perpendicular to $r$ equals to
\begin{equation}
v_\perp=\Omega r+\omega R\cos\gamma,\quad \omega=\sqrt{\frac{GM_1}{R^3}}
\label{vel}
\end{equation}
where $\omega$ is the disc angular velocity. From this, we can derive an $R$-dependent $\alpha$,
\begin{align}
&\alpha(R,\phi)= \left(\frac{M}{M_1}\right)^2\left(\frac{r}{a}\right)^2+ \left(\frac{M}{M_1}\right)^{3/2}\frac{r}{a} \left(\frac{R}{a}\right)^{-1/2}\cos\gamma,\label{f2_R}\\
&\alpha(R)=\frac{\int_0^{2\upi}\alpha(R,\phi){\rm d}\phi}{2\upi}.
\label{f2_disc}
\end{align}
A simple approximation to $\alpha(R)$ can be obtained by assuming the primary to be at rest, and calculating the specific angular momentum with respect to it. It yields,
\begin{equation}
\alpha(R)=\left(\frac{M}{M_1}\right)^2\left(\frac{R}{a}\right)^2+\left(\frac{M}{M_1}\right)^{3/2}\left(\frac{R}{a}\right)^{1/2}.
\label{f2_appr}
\end{equation}

If the outflow is from a close vicinity of the accretor, $R\ll a$. In this case, $r\rightarrow a_1$ and $r\cos\gamma\rightarrow R-a_1\cos\phi\rightarrow R$ when averaged over $\phi$. Equation (\ref{f2_disc}) then implies
\begin{equation}
\alpha(R\ll a)\simeq q^2 \ll 1,
\label{f2_est}
\end{equation}
which equals $\simeq 0.0036$ for $q=0.06$. We note that this value of $\alpha$ corresponds to the specific angular momentum of accretor. (On the other hand, the approximation \ref{f2_appr} gives an incorrect value of $\alpha(R\rightarrow 0)=0$.) Thus, an outflow from an inner part of the accretion disc (as in the case of the luminosity comparable to the Eddington one) carries a very small fraction of its original angular momentum and $\alpha\ll 1$. 

\begin{figure}
\centerline{\includegraphics[width=\columnwidth]{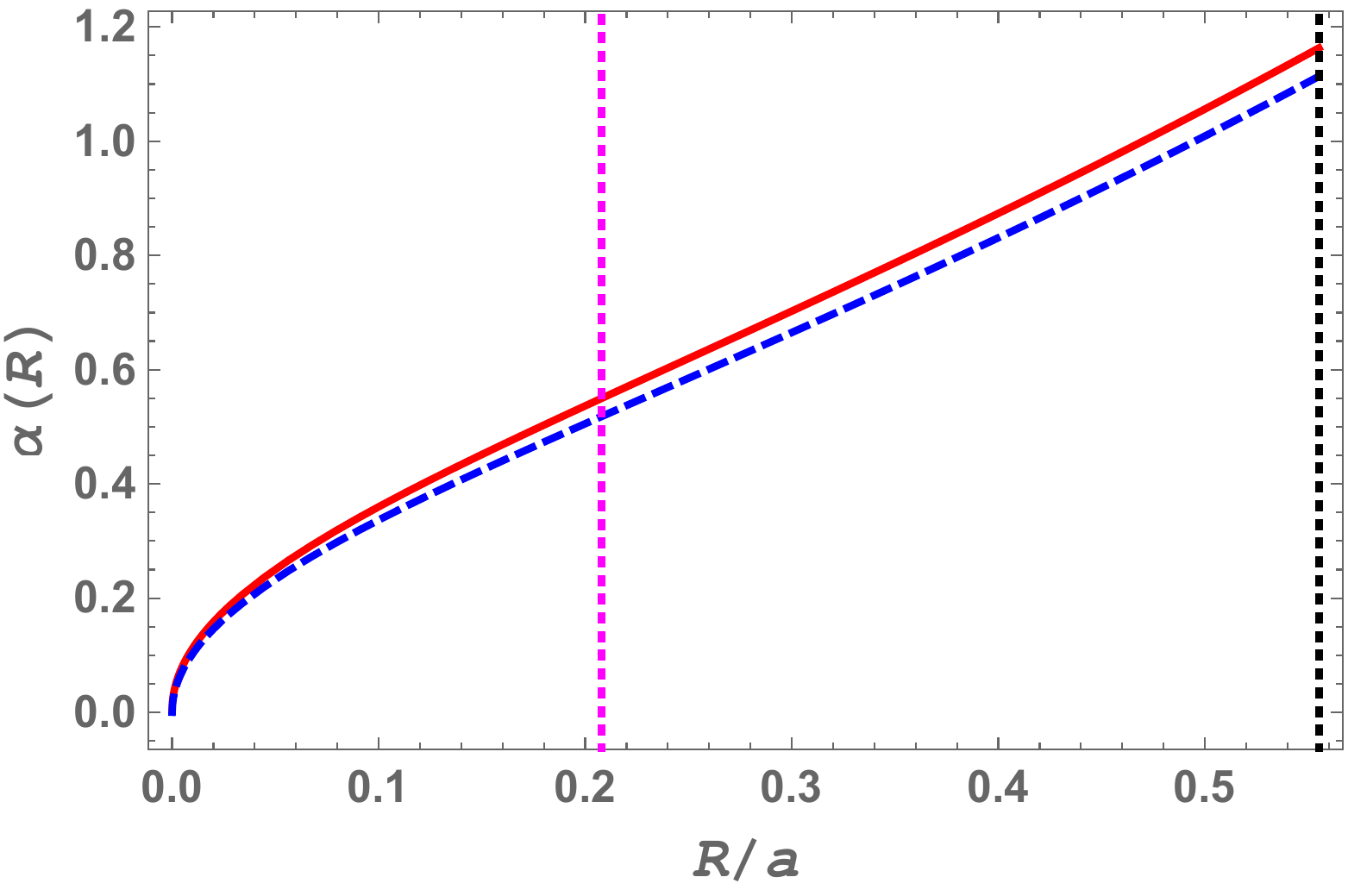}}
\caption{The dependence of the fraction of the specific angular momentum of the secondary carried away by a local outflow as a function of the distance from the compact object at which the outflow takes place for $q=0.06$. The solid red curve shows the exact dependence of equation (\ref{f2_disc}), and the dashed blue curve shows the approximation of equation (\ref{f2_appr}). The vertical magenta (at $R\simeq 0.21 a$) and black (at $R=0.9 R_1\simeq 0.55 a$) dotted lines show the circularization radius and the estimate of the tidal radius, respectively.
}
   \label{f2_plot}
  \end{figure}

The dependence of $\alpha$ on $R/a$ for $q=0.06$ is shown in Fig.\ \ref{f2_plot}. We find that $\alpha\simeq \alpha_{L1}$ at the circularization radius of $R_{\rm circ}\simeq 0.21 a\simeq 0.34 R_1$, where $R_1\simeq 0.62 a$ \citep{eggleton83}. On the other hand, the disc can extend to the tidal radius, around which $\alpha\sim 1$. Thus, a disc wind from an outer part of the disc can carry away even most of the specific angular momentum flowing from the donor. Fig.\ \ref{f2_plot} also shows the approximation to $\alpha(R)$ of equation (\ref{f2_appr}). We see that while it gives an incorrect value for $R\rightarrow 0$, it is relatively accurate at larger $R$. 

\section{The evolutionary model}
\label{model}

\subsection{The method}
\label{method}

In order to calculate evolution of stripped giants, we use the Warsaw stellar-evolution code described in \citet{ziolkowski05}. In \citet{z16}, the code was calibrated to reproduce the Sun at the solar age. This calibration resulted in the chemical composition of the H mass fraction of $X = 0.74$, the metallicity of $Z = 0.014$, and the mixing length parameter of $\alpha = 1.55$.

Our program calculates evolutionary sequences of stripped giants for an assumed constant total mass, $M_2$, and a varying (growing) mass of the He core. The program calculates not only the radius and the luminosity of each model but also its entire internal structure. This structure is essential for calculating the reaction of the star to mass transfer. In the calculations, we followed the approach used to calculate models of GRS1915+105/V1487 Aql \citep{zz17} and IGR J17451--3022 \citep{z16}. The situation is now simpler than in those cases since V404 Cyg has the reliable estimate of the donor mass, see Section \ref{intro}. Therefore, we need to calculate the sequences of models for only limited range of the values of $M_2$.

\subsection{The core mass-radius and radius-luminosity planes}
\label{planes}

\begin{figure}
\centerline{\includegraphics[width=\columnwidth]{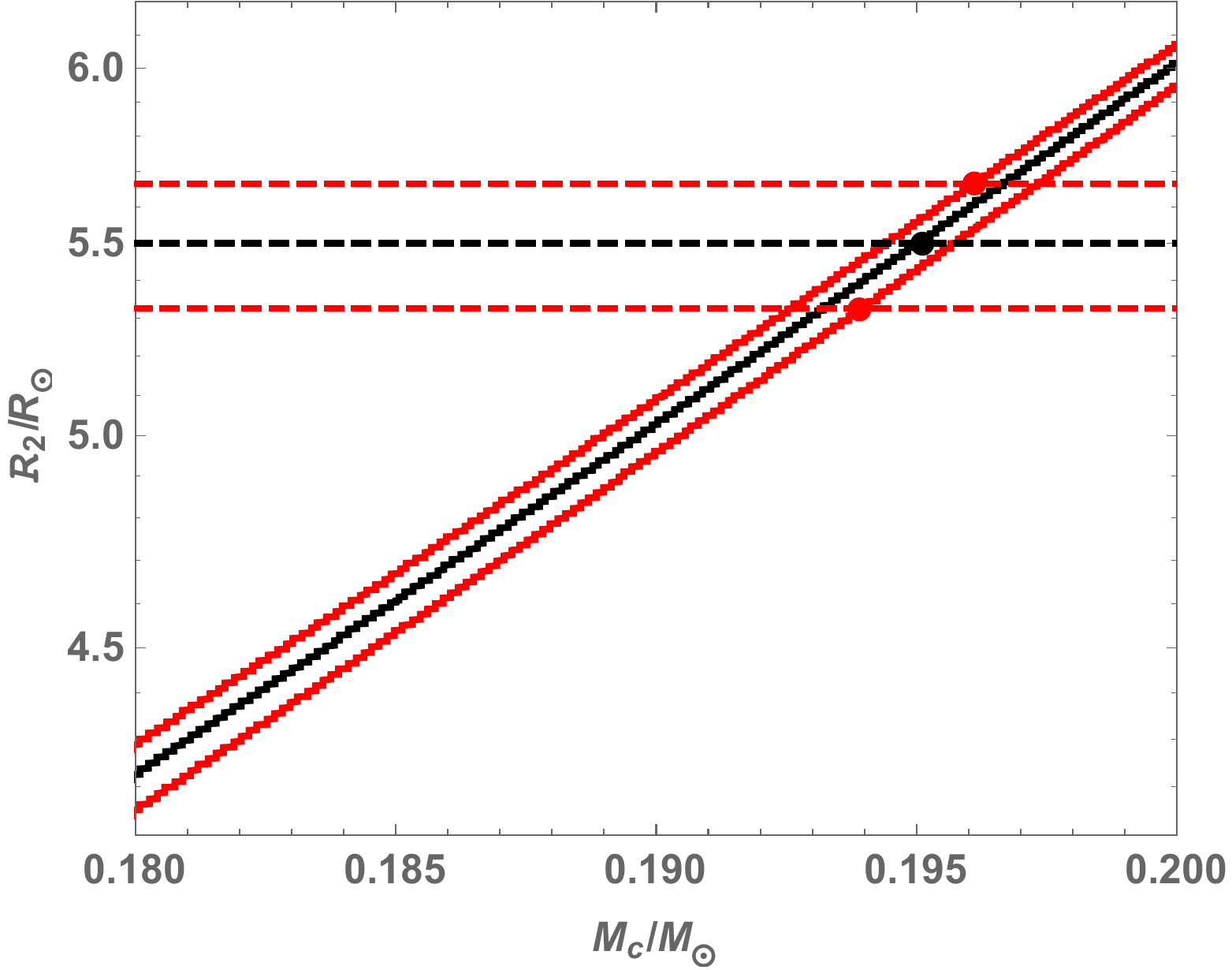}}
\caption{Evolution of partially stripped giants for $M_2=0.54\msun$ (black middle solid line), 0.49 (lower red solid line) and $0.59 \msun$ (upper lower red solid line) in the $M_{\rm c}$-$R_2$ diagram. The evolution proceeds (from left to right) at the constant total mass. The horizontal dashed lines show the radii of the Roche lobe around the donor (and so, to a good approximation, the radii of the star) for the above masses of the donor (with $R_2$ increasing with $M_2$). The black middle circle shows the position of our model for $M_2=0.54\msun$, the lower red cirle shows it for $M_2=0.49\msun$, and the upper red solid circle shows it for $0.59\msun$.}
   \label{evolution}
  \end{figure}

The results of our evolutionary calculations are presented in Fig.\ \ref{evolution}, which shows the evolutionary tracks for the stellar masses of 0.49, 0.54 and $0.59\msun$ of stripped giants in the core mass, $M_{\rm c}$, vs. $R_2$ plane. The stars evolve at constant mass and the driving mechanism is the progress of the H-burning shell moving outwards. The radii of the partially stripped giants generally increase with $M_{\rm c}$, except for the shrinking when the masses of their envelopes become very low (not shown). From crossings of these tracks with the corresponding lines showing the radii of the donor Roche lobe (equation \ref{roche}), we find the core mass of $M_{\rm c}\simeq 0.195\pm 0.001 \msun$.

\begin{figure}
\centerline{\includegraphics[width=\columnwidth]{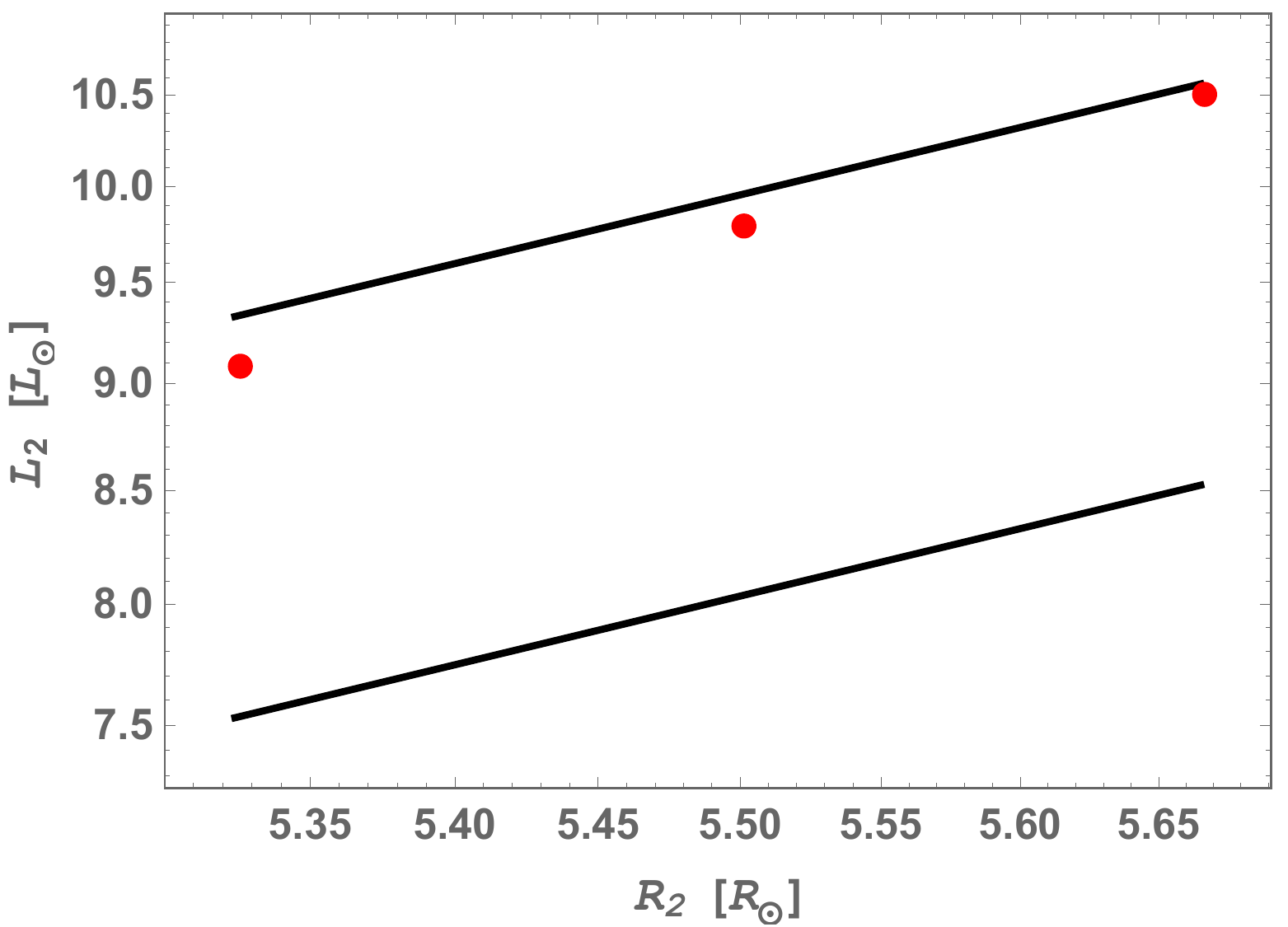}}
   \caption{A comparison of the luminosity predicted by our models (red circles) with the observational constraints for the allowed range of radii (lower and upper black lines). The middle red circle shows the position of our model for $M_2=0.54\msun$, and the other two correspond to $M_2=0.49$ (left) and $0.59\msun$ (right).}
   \label{luminosity}
  \end{figure}

Our evolutionary model predicts also the luminosity, $L_2$, as a function of $M_{\rm c}$, resulting from H burning in the shell. We compare the range predicted for the determined range of $M_{\rm c}$ with the range allowed observationally (see Section \ref{intro}). A comparison for $L_2(R_2)$ is given in Fig.\ \ref{luminosity}. We see that the agreement of the predicted luminosity as a function of the radius with the constraints derived from the stellar spectral type is very good. The fact that we reproduce quite precisely the luminosity of the donor gives support to both the correctness of our model and to the accuracy of the observational determination of the mass (implying the radius) and the temperature of the star.

On the other hand, the fitting formula (4) of \citet{webbink83} substantially underestimates the luminosity, yielding $4.4\lsun$ for our value of $M_{\rm c}=0.195\msun$. This value would then disagree with the observational constraints on the luminosity. This difference is due to the updated chemical compositions and more contemporary physics (especially opacities) used by us.

\section{Properties of the mass transfer in V404 Cyg}
\label{transfer}

\subsection{Constraints on the donor mass loss rate and the outflow parameters}
\label{constraints}

With the above results, we can calculate the rate of the mass transfer under different assumptions about the outflow. We assume our model for the current best determination of the donor mass, $0.54\msun$. Due to the growing mass of the core, the star will expand at the rate ${\rm d}R_2/{\rm d}t$, which will lead to the mass flow through the inner Lagrangian point. The rate of this flow will depend on the reaction of the radius of the Roche lobe around the donor to the mass transfer, and the process will be self-adjusting assuring that changes of the donor radius follow the changes of the radius of the Roche lobe. For a given rate of the mass outflow, there will be a dependence of the radius on the mass, ${\rm d}R_2/{\rm d}M_2=({\rm d}R_2/{\rm d}t)/\dot M_2$, which we need to equal to the corresponding dependence for the Roche lobe evolution, given by equation (\ref{rate}). 

\begin{figure}
\centerline{\includegraphics[width=\columnwidth]{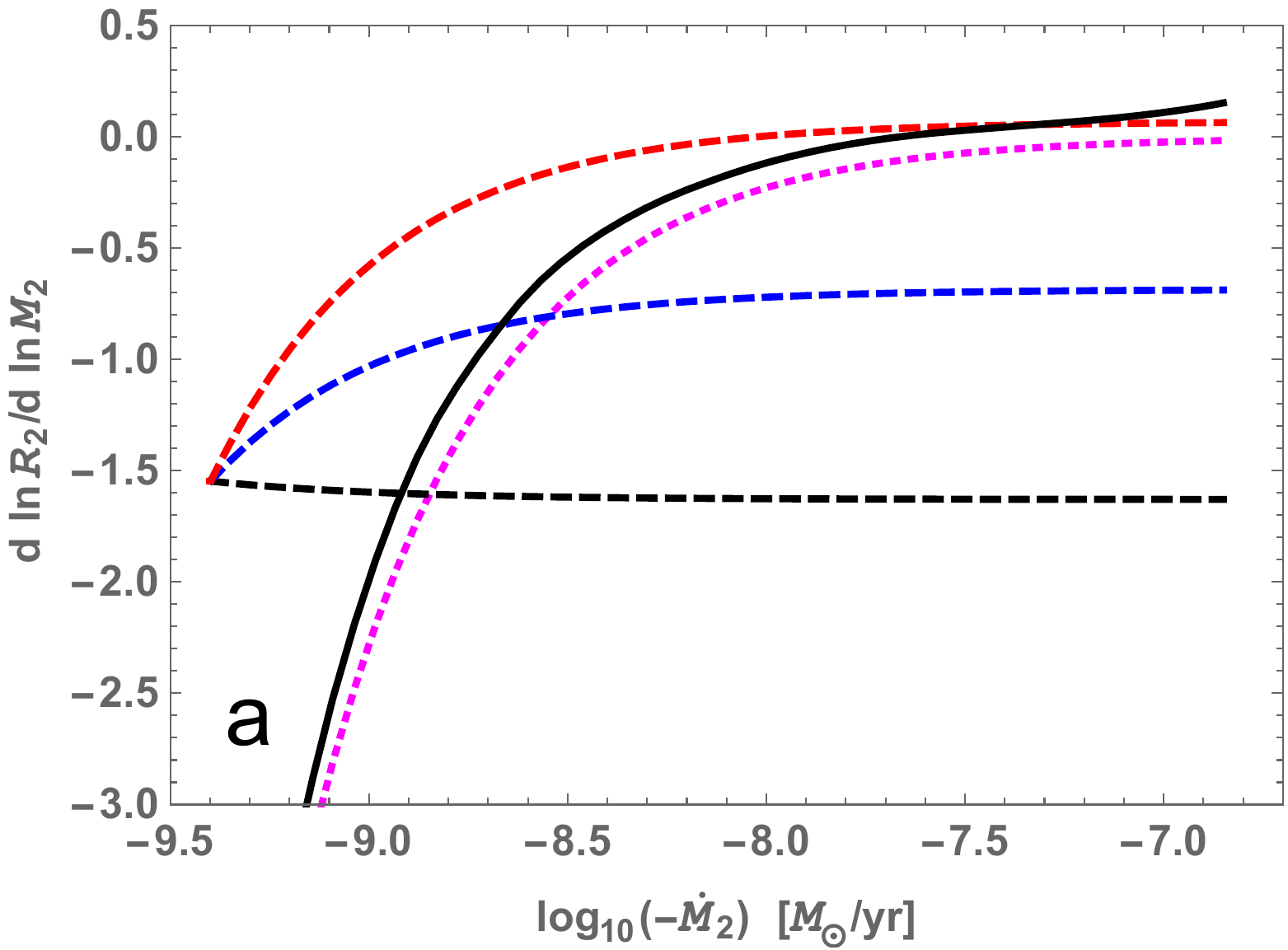}}
\centerline{\includegraphics[width=\columnwidth]{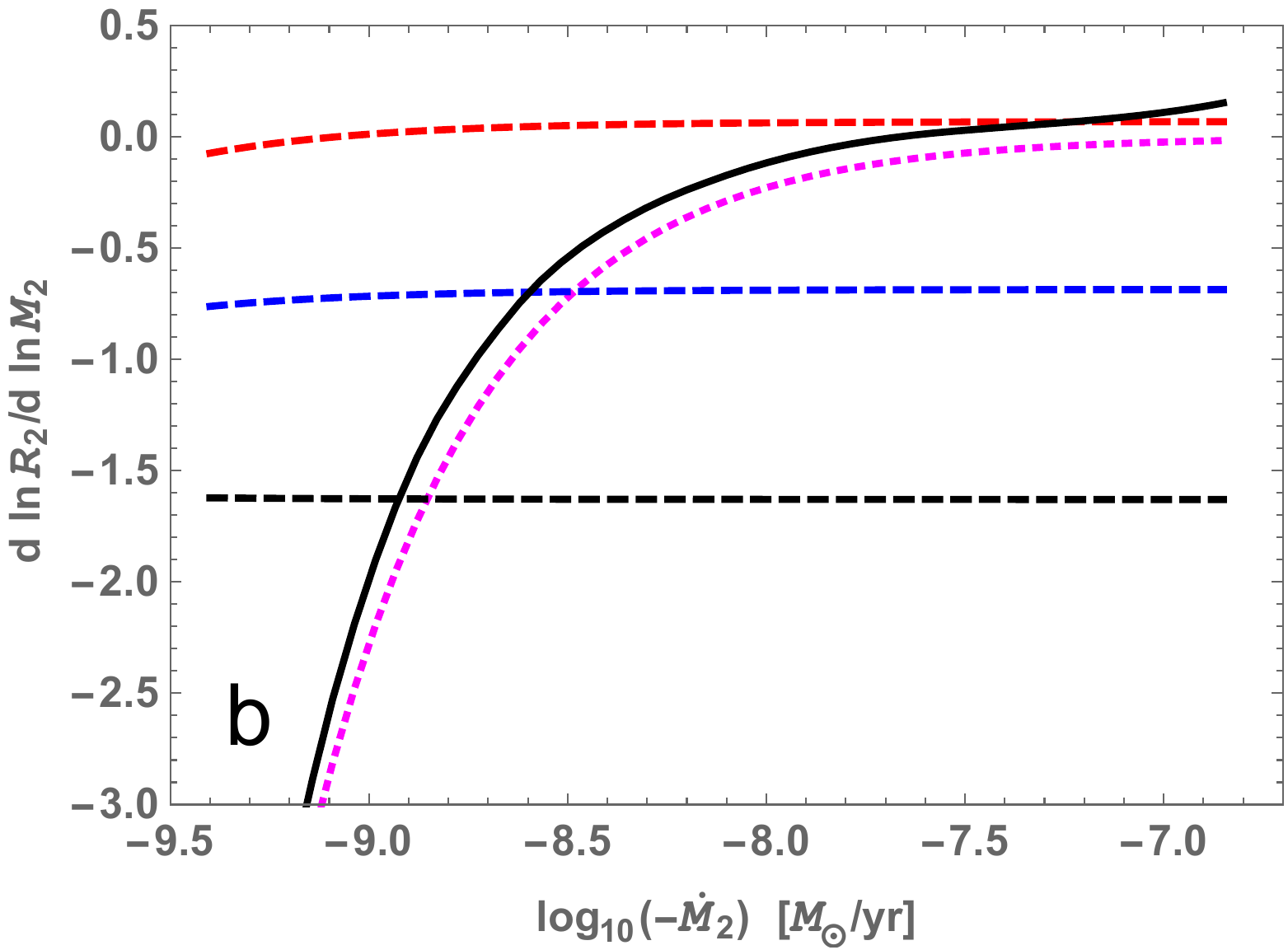}}
   \caption{The rates of the evolutionary change of the radius of the donor and that of its Roche lobe as functions of the donor mass loss rate, $-\dot M_2$. The panels (a) and (b) are for $\dot M_1= 4\times 10^{-10}\msun$ yr$^{-1}$ and $3.5\times 10^{-11}\msun$ yr$^{-1}$, respectively. The black solid curves give the results of our evolutionary model including the effect of mass removal from the outer layers of the star. The magenta dotted curves show the corresponding model of \citet{webbink83} (neglecting the mass removal). The rates of the Roche lobe change, ${\rm d}\ln R_2/{\rm d}\ln M_2$, are shown for $\alpha=0$ (bottom black dashed curves), 0.5 (middle blue dashed curves) and 0.9 (top red dashed curves). The values of $\beta$ along those curves are given by $\dot M_1/|\dot M_2|$. For (a) and (b) $\beta=1$, $\simeq$0.088 at the curve beginnings and $\beta\simeq 0.003$, 0.00018 at their ends, respectively. The intersections of the black solid curves and the dashed curves give possible self-consistent models of \source, see text for details.
}
\label{derivative}
  \end{figure}

For low rates of the mass outflow, the evolutionary expansion dominates and the effect of the mass removal from the outer layers of the star is minor. Then ${\rm d}R_2/{\rm d}M_2<0$, i.e., the star expands. However, for sufficiently high outflow rates, the fast mass removal from the outer layers leads to a non-equilibrium configuration, and the star starts to shrink, ${\rm d}R_2/{\rm d}M_2>0$. The logarithmic derivative ${\rm d}\ln R_2/{\rm d}\ln M_2=(M_2/R_2) {\rm d}R_2/{\rm d}M_2$ for our model with $M_2= 0.54\msun$ is shown by the black solid curves in Figs.\ \ref{derivative}(a,b). We see that the star starts to shrink at $\dot M_2\ga 2\times 10^{-8}\msun$ yr$^{-1}$. 

We can compare our results with those of \citet{webbink83}, who took into account only the evolutionary growth of the star (see their equation 13); thus their value of ${\rm d}R_2/{\rm d}t$ is independent of $M_2$. Then, their time derivative for the considered model depends only on the core mass and the rate of the core mass growth (given by the luminosity and the H content in the H-burning shell, $X$), see their equation (14). That dependence for our best-fit values of $M_2=0.54\msun$, $L_2=8.9\lsun$, $M_{\rm c}= 0.195\msun$, and $X=0.54$ at the present evolution stage (as found by our model) is plotted in the magenta dotted curve in Figs.\ \ref{derivative}(a,b), and it is given by,
\begin{equation}
\frac{{\rm d}\ln R_2}{{\rm d}\ln M_2} \simeq -\frac{2.28\times 10^{-9}\msun\,{\rm yr}^{-1}}{-\dot M_2}.
\end{equation}
We see that the dependence of \citet{webbink83} is somewhat lower than ours. Also, it predicts the radius derivative to be always negative, while the effect of mass removal from the outer layers of the star becomes important at high values of $-\dot M_2$, and the derivative becomes positive, as found in our solution.

The evolutionary change of the radius equals that of the Roche lobe change, given by equation (\ref{rate}). The latter depends on the outflow parameters $\beta$ and $\alpha$. Following our review of the estimates of the average accretion rate in \source\/ in Section \ref{intro}, we adopt fiducial values of $\dot M_1=\langle \dot M_1\rangle =4\times 10^{-10}\msun$ yr$^{-1}$ and $3.5\times 10^{-11}\msun$ yr$^{-1}$, though we consider the latter value less likely. While the actual value of $\dot M_1$ remains relatively uncertain, the former estimate represents, most likely, a firm upper limit to $\dot M_1$, see Section \ref{intro}. In Figs.\ \ref{derivative}(a,b), we plot ${\rm d}\ln R_2/{\rm d}\ln M_2(\beta,\alpha)$ (where $\beta\equiv \dot M_1/|\dot M_2|$) as a function of $-\dot M_2$ for the values of $\alpha=0$, 0.5 and 0.9.

Self-consistent solutions have to lie on the intersections of the curve describing our evolutionary model with those for the Roche lobe. The value of $\beta$ follows then from the value of $-\dot M_2$ at that intersection. Analysing Figs.\ \ref{derivative}(a) and (b), we find that at $\alpha= 0$, $-\dot M_2\simeq 1.2\times 10^{-9}\msun$ yr$^{-1}$ in both considered cases, implying $\beta\la 0.33$, 0.03, respectively, for $\alpha\geq 0$, i.e., the mass transfer has to be highly non-conservative. For $\alpha= 0.5$, $-\dot M_2\simeq 2.2$ and $2.5\times 10^{-9}\msun$ yr$^{-1}$ and $\beta\simeq 0.19$, 0.014, and for $\alpha= 0.9$, $-\dot M_2\simeq 4.8$ and $5.9\times 10^{-8}\msun$ yr$^{-1}$ and $\beta\simeq 0.008$, 0.0006, for $\dot M_1=4\times 10^{-10}$ and $3.5\times 10^{-11}\msun$ yr$^{-1}$, respectively. Those solutions imply that almost all transferred mass has to leave the binary. Solutions for other values of $\alpha$ can be readily found by plotting further dependencies of ${\rm d}\ln R_2/{\rm d}\ln M_2(\beta,\alpha)$ on Fig.\ \ref{derivative} and determining their intersection with the black solid curve.

We note that at the solutions corresponding to $\alpha=0.9$, the donor does not remain in thermal equilibrium anymore, and therefore it is able to shrink. However, the solution is nevertheless dynamically stable since the evolutionary calculations indicate that the donor evolving on thermal time scale is able to adjust its radius to that of the shrinking Roche lobe. \citet{webbink83} emphasized that they consider only donors evolving on nuclear time scale and therefore their models can only expand, see the magenta curves in Figs.\ \ref{derivative}(a,b).

As discussed in the Section \ref{outflow}, outflows from the vicinity of the accretor (in the form of strongly ionized winds and jets) have $\alpha\simeq 0$, which corresponds to our first solution above. On the other hand, outflows from outer parts of the accretion disc (in the form of neutral winds; \citealt{munoz_darias16}), will have an intermediate value of $\alpha$, and may roughly correspond to our solution with $\alpha=0.5$. We note that \citet{munoz_darias16} have apparently been unable to estimate the range of radii of the neutral outflows they found. Solutions with $\alpha\ga 0.9$ correspond to outflows from near the outer edge of the disc and are also in principle possible, see Section \ref{outflow}. However, they require both the loss of almost all matter flowing from the donor and fine-tuning of the donor paramers; we thus consider them less likely for \source\ than the other solutions.

The main caveat to our results concerns the estimate of the mass accreted during the outbursts on the observed X-ray light curves. This requires the knowledge of the absorption (appearing to be strong) during outbursts and the radiative efficiency, with both being uncertain. Still, our adopted value most likely represents a firm upper limit. Furthermore, it appears that only a small fraction of the total disc mass is accreted during an outburst \citep{zycki99}, and then it is uncertain whether the mass accreted during an outburst is directly related to that supplied to the disc during the preceding period of quiescence. Nevertheless, an equality of the two masses is expected when averaged over many outbursts.

\subsection{Implication for the evolution of the orbital period}
\label{period}

Another potential diagnostic of the presence of an outflow in \source\ would be a measurement of a change of its orbital period, since the values of the parameters characterising the outflow influence the expected changes of $P$. This diagnostic was utilized by \citet{disalvo08}, \citet{burderi09,burderi10} and \citet{marino17} to constrain the parameters of the mass outflow from the binaries they considered. The relation between the orbital period change, the rate of mass flow from the donor, and the parameters $\alpha$ and $\beta$ is given by
\begin{equation}
\frac{\dot P}{P} = -\frac{\dot M_2}{M_2} \left[3 - 3\beta\frac{M_2}{M_1} -(1-\beta)\frac{M_2}{M} -3(1-\beta)\alpha\frac{M_1}{M}\right].
\label{Pdot}
\end{equation}
At $\beta=1$, $\alpha=0$, this reduces to the standard expression for $\dot P$ for conservative mass transfer (e.g., \citealt{frank02}), 
\begin{equation}
\frac{\dot P}{P} = -3\dot M_2\left(\frac{1}{M_2}-\frac{1}{M_1}\right), 
\label{standard}
\end{equation}
implying an increasing period for mass transfer from the less massive to the more massive star. We find that $\dot P>0$ remains the case generally for any values of $\beta$ and $\alpha$. In the case of extreme non-conservative transfer, $\beta=0$, $\alpha=1$, we have 
\begin{equation}
\frac{\dot P}{P} = -2\frac{\dot M_2}{M}, 
\label{extreme}
\end{equation}
which applies to the donor mass loss by stellar wind, see \citet{davidsen74}. We can see that, for a given $\dot M_2$, the presence of an outflow leads to a decrease of $\dot P$ with respect to the conservative case in most cases. However, it leads to a substantial increase of $\dot P$ for a given $\dot M_1$, since $\dot P\propto -\dot M_2 \propto \dot M_1/\beta$. A more general form of equation (\ref{Pdot}), namely taking into account the effect of the gravitational radiation, is given by equation (5) of \citet{disalvo08}.

We have calculated the expected changes of the orbital period for all six
solutions shown in Fig.\ \ref{derivative}. At $\alpha=0$, $\dot P\simeq 1.15\times 10^{-10}$, and the time scale of the $\dot P$ increase is $P/\dot P\simeq 1.5\times 10^8$ yr for both considered values of $\dot M_2$. At $\alpha=0.5$, the time scale is very similar, $\simeq 1.4\times 10^8$ yr in both cases. Almost the same time scale is found for conservative mass transfer and $-\dot M_2=\dot M_1=1.2\times 10^{-9}\msun$ yr$^{-1}$, which would be the case if we underestimated the actual mass accretion rate. The time scale becomes much shorter for $\alpha=0.9$ (which case, however, we consider less likely than lower values of $\alpha$), $\simeq 2.7$ and $2.3\times 10^7$ yr for $\dot M_1 =4\times 10^{-10}$ and $3.5\times 10^{-11}\msun$ yr$^{-1}$, respectively. Unfortunately, no information about orbital period changes of \source\ is available so far, and it may be difficult to obtain given its relatively long orbital period.

\section{Conclusions}

Our main results can be summarized as follows.

We have considered non-conservative mass transfer in a binary, which is parameterized by the fractions of the mass and angular momentum leaving the system. We have calculated the latter parameter, $\alpha$, as a function of the distance from the accretor along its accretion disc. 

We have estimated the mass accretion rate in \source\ averaged over intervals between its outbursts. While its value remains relatively uncertain, the best available upper limit appears to be $\langle \dot M_1\rangle =4\times 10^{-10}\msun$ yr$^{-1}$. This value can be compared to that estimated theoretically given that the donor is a giant with the well-established mass and spectral type (which parameters yield the luminosity in good agreement with the theoretical value from our evolutionary model). From those estimates, there is a significant discrepancy of that $\langle \dot M_1\rangle$ with the donor mass loss rate determined based on evolutionary models, $-\dot M_2$, with the former being at least three times lower. 

This discrepancy can be resolved taking into account the likely presence of outflows during the outbursts. We find that at least 70 per cent of the mass lost from the donor has to leave the binary. The allowed solution for the actual donor mass loss rate is parameterized by our two outflow parameters. Our results are in agreement with either the observed outflows from the outer disc \citep{munoz_darias16} or with the character of accretion observed during the 2015 outburst \citep{motta17b}, compatible with outflows in the vicinity of the black hole. In the latter case, the matter reaching that region had lost most of its angular momentum before outflowing. In the former case, there is no available estimate of the disc radius from which outflows originate, and an intermediate value of $\alpha$ appears appropriate, which we illustrate by the case with $\alpha=0.5$.

We also calculate the expected rate of the orbital period increase, parameterized by the donor mass loss rate and the two outflow parameters. We find a characteristic increase time scale of $\sim\! 10^8$ yr. 

Finally, we improve the accuracy of the standard approximation for the location of the $L_1$ point (Appendix \ref{L1_app}), which was divergent at low mass ratios of $q\rightarrow 0$. Our approximation is no more divergent, and its accuracy is $\la 3$ per cent for almost all mass ratios. 

\section*{Acknowledgements}

We thank the referee, Luciano Burderi, and Tiziana Di Salvo for valuable comments, and Piotr \.Zycki for discussions. This research has been supported in part by the Polish National Science Centre grants 2013/10/M/ST9/00729, 2015/18/A/ST9/00746 and 2016/21/P/ST9/04025.

\appendix
\section{The location of $\mathbfit{L}_\mathbf{1}$}
\label{L1_app}

\begin{figure}
\centerline{\includegraphics[width=\columnwidth]{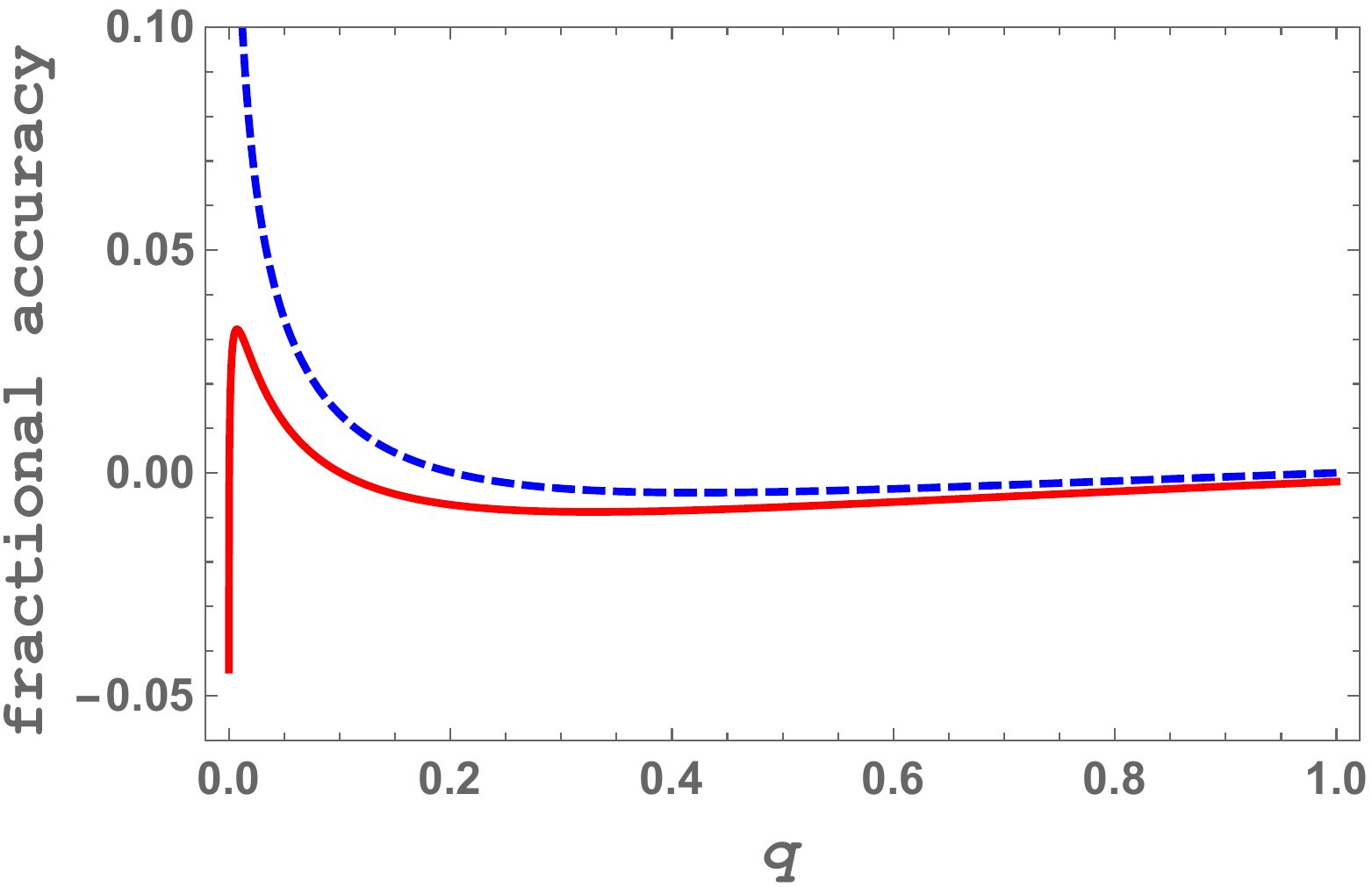}}
\caption{The fractional accuracy of the approximation to the distance between $L_1$ and the centre of the primary of equation (\ref{plavec2}), shown by the red solid curve, and that of the original approximation of \citet{frank02}, the blue dashed curve.}
   \label{L1_plot}
  \end{figure}

The distance of the $L_1$ point to the centre of the primary, $b_1$, is given by the real root of, e.g., equation (2.23) of \citet{kopal78}. \citet{frank02} give the following approximation to it,
\begin{equation}
\frac{b_1(q)}{a}\simeq \frac{1}{2}-0.227 \log_{10} q,
\label{plavec}
\end{equation}
citing an unpublished work of Plavec \& Kratochvil. However, we note that $b_1(q)\rightarrow\infty$ at $q\rightarrow 0$, and this approximation exceeds the accurate result by more than 7 per cent already for $q\leq 0.02$. We propose its simple modification, 
\begin{equation}
\frac{b_1(q)}{a}\simeq \frac{1}{2}-0.227 \log_{10}(q+0.01),
\label{plavec2}
\end{equation}
which accuracy is $\la$3 per cent for $q\geq 10^{-5}$ (and it underestimates $b_1=a$ at $q=0$ by 4.6 per cent). The fractional accuracy, $b_1({\rm approximation})/b_1({\rm exact})-1$, of both approximations is shown in Fig.\ \ref{L1_plot}. We note that our approximation should be used for $q\leq 1$ only, consistent with its definition as the distance to the primary. For $q>1$, $M_2$ becomes the primary, and equation (\ref{plavec2}) with $1/q$ as the argument gives then the distance to the centre of $M_2$. 

\label{lastpage}

\begin{thebibliography}{}

\bibitem[\protect\citeauthoryear{Barthelmy et al.}{2015}]{barthelmy15}
Barthelmy S. D., et al., 2015, GCN, 17929

\bibitem[\protect\citeauthoryear{Beardmore et al.}{2016}]{beardmore16} Beardmore A.~P., Willingale R., Kuulkers E., Altamirano D., Motta S.~E., Osborne J.~P., Page K.~L., Sivakoff G.~R., 2016, MNRAS, 462, 1847 

\bibitem[\protect\citeauthoryear{Burderi et al.}{2009}]{burderi09} Burderi L., Riggio A., Di Salvo T., Papitto A., Menna M.~T., D'A{\`i} A., Iaria R., 2009, A\&A, 496, L17 

\bibitem[\protect\citeauthoryear{Burderi et al.}{2010}]{burderi10} Burderi L., Di Salvo T., Riggio A., Papitto A., Iaria R., D'A{\`i} A., Menna M.~T., 2010, A\&A, 515, A44 

\bibitem[\protect\citeauthoryear{Cox}{2000}]{cox00}
Cox, A. N., 2000, Allen's astrophysical quantities, 4th ed., Springer

\bibitem[\protect\citeauthoryear{Casares \& Charles}{1994}]{cc94}
Casares J., Charles P.A., 1994, MNRAS, 271, L5

\bibitem[\protect\citeauthoryear{Casares, Charles \& Naylor}{Casares et al.}{1992}]{casares92}
Casares J., Charles P. A., Naylor T., 1992, Nature, 355, 614

\bibitem[\protect\citeauthoryear{Charles et al.}{1989}]{charles89} Charles P.~A., et al., 1989, in The 23rd ESLAB Symposium on Two Topics in X Ray Astronomy, ESA, p.\ 103

\bibitem[\protect\citeauthoryear{Chen, Shrader \& Livio}{Chen et al.}{1997}]{chen97} Chen W., Shrader C.~R., Livio M., 1997, ApJ, 491, 312 

\bibitem[\protect\citeauthoryear{Davidsen \& Ostriker}{1974}]{davidsen74} 
Davidsen A., Ostriker J.~P., 1974, ApJ, 189, 331 

\bibitem[\protect\citeauthoryear{Di Salvo et al.}{2008}]{disalvo08} Di Salvo T., Burderi L., Riggio A., Papitto A., Menna M.~T., 2008, MNRAS, 389, 1851 

\bibitem[\protect\citeauthoryear{Eggleton}{1983}]{eggleton83} Eggleton P.~P., 1983, ApJ, 268, 368 

\bibitem[\protect\citeauthoryear{Frank, King \& Raine}{Frank et al.}{2002}]{frank02} Frank J., King A., Raine D.~J., 2002, Accretion Power in Astrophysics, Cambridge University Press 

\bibitem[\protect\citeauthoryear{Iaria et al.}{2018}]{iaria18} Iaria R., et al., 2018, MNRAS, 473, 3490 

\bibitem[\protect\citeauthoryear{Kimura et al.}{2016}]{kimura16}
Kimura M. et al., 2016, Nature, 529, 54

\bibitem[\protect\citeauthoryear{King}{1993}]{king93}
King, A.R., 1993, MNRAS, 260, L5

\bibitem[\protect\citeauthoryear{Khargharia, Froning \& Robinson}{Khargharia et al.}{2010}]{khargharia10}
Khargharia J., Froning C. S., Robinson E. L., 2010, ApJ, 716, 1105

\bibitem[\protect\citeauthoryear{Kopal}{1978}]{kopal78} Kopal Z., 1978, Dynamics of close binary systems, Astrophysics and Space Science Library, 68. Dordrecht: Reidel 

\bibitem[\protect\citeauthoryear{Makino}{1989}]{makino89}
 Makino F., 1989, IAUC, 4782, 1

\bibitem[\protect\citeauthoryear{Marino et al.}{2017}]{marino17} Marino A., Di Salvo T., Gambino A.~F., Iaria R., Burderi L., Matranga M., Sanna A., Riggio A., 2017, A\&A, 603, A137 

\bibitem[\protect\citeauthoryear{Mart{\'{i}}, Luque-Escamilla and Garc{\'{i}}a-Hern{\'{a}}ndez}{Mart\'{i} et al.}{2016}]{marti16}
Mart{\'{i}} J., Luque-Escamilla P. L., Garc{\'{i}}a-Hern{\'{a}}ndez M. T., 2016, A\&A, 586, 58

 \bibitem[\protect\citeauthoryear{Miller-Jones et al.}{2009}]{miller_jones09}
 Miller-Jones J. C. A., et al., 2009, ApJ, 706, L230

\bibitem[\protect\citeauthoryear{Motta et al.}{2016}]{motta16}
 Motta S. E., et al., 2016, ATel, 8510

\bibitem[\protect\citeauthoryear{Motta et al.}{2017a}]{motta17a} Motta S.~E., et al., 2017a, MNRAS, 468, 981 

\bibitem[\protect\citeauthoryear{Motta et al.}{2017b}]{motta17b} Motta S.~E., et al., 2017b, MNRAS, 471, 1797 

\bibitem[\protect\citeauthoryear{M{\~u}noz-Darias et al.}{2016}]{munoz_darias16}
M{\~u}noz-Darias T., et al., 2016, Nature, 534, 75

\bibitem[\protect\citeauthoryear{Paczy\'nski}{1967}]{paczynski67}
 Paczy\'nski B., 1967, Acta Astron., 17, 287

\bibitem[\protect\citeauthoryear{Paczy\'nski \& Zi\'o{\l}kowski}{1967}]{pz67}
Paczy\'nski B., Zi\'o{\l}kowski, J., 1967, Acta Astron., 17, 7

\bibitem[\protect\citeauthoryear{Poutanen et al.}{2007}]{poutanen07} Poutanen J., Lipunova G., Fabrika S., Butkevich A.~G., Abolmasov P., 2007, MNRAS, 377, 1187 

\bibitem[\protect\citeauthoryear{Rappaport, Joss \& Webbink}{Rappaport et al.}{1982}]{rappaport82} Rappaport S., Joss P.~C., Webbink R.~F., 1982, ApJ, 254, 616 

\bibitem[\protect\citeauthoryear{Rodriguez et al.}{2015}]{rodriguez15}
 Rodriguez J., et al., 2015, A\&A, 581, L9

\bibitem[\protect\citeauthoryear{S{\c a}dowski \& Narayan}{2015}]{sadowski15} S{\c a}dowski A., Narayan R., 2015, MNRAS, 453, 3213 

\bibitem[\protect\citeauthoryear{S{\'a}nchez-Fern{\'a}ndez et al.}{2017}]{sanchez17} S{\'a}nchez-Fern{\'a}ndez C., Kajava J.~J.~E., Motta S.~E., Kuulkers E., 2017, A\&A, 602, A40 

\bibitem[\protect\citeauthoryear{Sanna et al.}{2017}]{sanna17} Sanna A., et al., 2017, MNRAS, 471, 463 

\bibitem[\protect\citeauthoryear{Tout \& Hall}{1991}]{tout91} Tout C.~A., Hall D.~S., 1991, MNRAS, 253, 9 

\bibitem[\protect\citeauthoryear{Verbunt}{1993}]{verbunt93} Verbunt F., 1993, ARA\&A, 31, 93 

\bibitem[\protect\citeauthoryear{Wagner et al.}{1989}]{wagner89}
 Wagner R. M., Starrfield S. G., Cassatella A., Hurst G. M., Mobberley M., Marsden B. G., 1989, IAUC, 4783, 1

\bibitem[\protect\citeauthoryear{Webbink, Rappaport \& Savonije}{Webbink et al.}{1983}]{webbink83}
Webbink R. F., Rappaport, S. A., Savonije, G. J., 1983, ApJ, 270, 678

\bibitem[\protect\citeauthoryear{Yuan \& Narayan}{2014}]{yuan14} Yuan F., Narayan R., 2014, ARA\&A, 52, 529 

\bibitem[\protect\citeauthoryear{Zdziarski et al.}{2016}]{z16} 
Zdziarski A.~A., Zi{\'o}{\l}kowski J., Bozzo E., Pjanka P., 2016, A\&A, 595, A52 

\bibitem[\protect\citeauthoryear{Zi\'o{\l}kowski}{1985}]{ziolkowski85}
Zi\'o{\l}kowski J., 1985, Acta Astron., 35, 199

\bibitem[\protect\citeauthoryear{Zi\'o{\l}kowski}{2005}]{ziolkowski05}
Zi\'o{\l}kowski J., 2005, MNRAS, 358, 851

\bibitem[\protect\citeauthoryear{Zi\'o{\l}kowski \& Zdziarski}{2017}]{zz17}
 Zi\'o{\l}kowski J., Zdziarski A. A., 2017, MNRAS, 469, 3315

\bibitem[\protect\citeauthoryear{{\.Z}ycki, Done, \& Smith}{{\.Z}ycki et al.}{1999}]{zycki99} {\.Z}ycki P.~T., Done C., Smith D.~A., 1999, MNRAS, 309, 561 

\end{thebibliography}
\end{document}